\documentclass[a4paper]{jpconf}
\usepackage{amsmath,amssymb,mathrsfs}
\usepackage{psfrag}
\usepackage{graphicx}
\usepackage{graphics}
\usepackage{epsfig}
\usepackage{bm}
\usepackage{color}
\usepackage{verbatim,color,ulem}
\newcommand{\beq}{\begin{equation}}
\newcommand{\eeq}{\end{equation}}
\newcommand{\beqa}{\begin{eqnarray}}
\newcommand{\eeqa}{\end{eqnarray}}

\begin{document}

\title{Fermionic condensation in ultracold atoms, 
nuclear matter and neutron stars}

\author{Luca Salasnich}

\address{Dipartimento di Fisica e Astronomia ``Galileo Galilei'' and CNISM, 
Universit\`a di Padova, \\
Via Marzolo 8, 35131 Padova, Italy}

\ead{luca.salasnich@unipd.it}

\begin{abstract}
We investigate the Bose-Einstein condensation of fermionic pairs 
in three different superfluid systems: 
ultracold and dilute atomic gases, bulk neutron matter, 
and neutron stars. In the case of dilute gases made of fermionic atoms 
the average distance between atoms is much larger than 
the effective radius of the inter-atomic potential. Here 
the condensation of fermionic pairs is analyzed as a function 
of the s-wave scattering length, which can be tuned in 
experiments by using the technique of Feshbach resonances 
from a small and negative value (corresponding to 
the Bardeen-Cooper-Schrieffer (BCS) regime of Cooper Fermi pairs) 
to a small and positive value (corresponding to the regime of 
the Bose-Einstein condensate (BEC) of molecular dimers), crossing the 
unitarity regime where the scattering length diverges.  
In the case of bulk neutron matter 
the s-wave scattering length of neutron-neutron potential 
is negative but fixed, and the condensate fraction of neutron-neutron 
pairs is studied as a function of the total neutron density. 
Our results clearly show a BCS-quasiunitary-BCS crossover 
by increasing the neutron density. Finally, in the case of 
neutron stars, where again the neutron-neutron scattering length is 
negative and fixed, we determine the condensate fraction 
as a function of the distance from the center of the neutron star, 
finding that the maximum condensate fraction appears 
in the crust of the neutron star. 
\end{abstract}

\section{Introduction}

In 1951 Penrose and Onsager \cite{penrose} introduced 
the idea of off-diagonal long-range order (ODLRO) 
of the one-body density matrix 
to determine the Bose-Einstein condensate fraction 
in a system of interacting bosons. 
In 1962 Yang \cite{yang} proved that for attractive 
fermions the Bose-Einstein condensation of fermionic pairs is instead 
related to the ODLRO of the two-body density matrix. 

It is now established that at zero temperature 
the Bose-Einstein condensate fraction of bosonic liquid $^4$He is below $10\%$ 
\cite{ceperly}, while 
for dilute and ultracold bosonic alkali-metal atoms it can 
reach $100\%$ \cite{review}. Some years ago, by using the ODLRO 
of the two-body density matrix, Salasnich, Manini 
and Parola \cite{sala-odlro} calculated 
the condensate fraction of fermionic pairs in the crossover from the 
Bardeen-Cooper-Schrieffer (BCS) 
state of Cooper Fermi pairs to the Bose-Einstein condensate (BEC) 
of molecular dimers at zero temperature (later in the 
same year there were other two papers \cite{ortiz,astra} on the 
same topic). In particular, 
it was found that the condensate fraction grows 
from zero to one in the BCS-BEC crossover \cite{sala-odlro,ortiz,astra}. 
These theoretical predictions are in quite good agreement 
with the data obtained in two experiments \cite{zwierlein,ueda} 
with Fermi vapors of $^6$Li atoms. 
Recently, the condensate fraction of bulk neutron matter 
has been calculated by Wlazlowski and Magierski \cite{magierski1,magierski2} 
and also by Salasnich \cite{sala-neutron}. 

In this paper we review the zero-temperature mean-field theory 
we have used to extract the fermionic 
condensate fraction in ultracold atoms \cite{sala-odlro} 
and bulk neutron matter \cite{sala-neutron}. In addition 
we determine the condensate fraction of neutron-neutron pairs 
as a function of the distance from the center of 
a neutron star \cite{star,zane}. In particular, we find the maximum 
condensate fraction at the distance $r/R\simeq 0.96$, 
where $R$ is the star radius.

\section{Bosonic and fermionic condensation} 

A quantum system of interacting identical bosons can be described 
by the bosonic field operator ${\hat \phi}({\bf r})$, which satisfies 
the familiar commutation rules \cite{leggett}
\beq 
\left[ 
{\hat \phi}({\bf r}), {\hat \phi}^+({\bf r}')
\right] = \delta({\bf r}-{\bf r}') \; , 
\quad\quad 
\left[
{\hat \phi}({\bf r}), {\hat \phi}({\bf r}')
\right] = 
\left[
{\hat \phi}^+({\bf r}), {\hat \phi}^+({\bf r}')
\right] = 0 \; , 
\eeq
where $\left[{\hat A},{\hat B}\right]={\hat A}{\hat B}-{\hat B}{\hat A}$ and 
$\delta({\bf r})$ is the Dirac delta function. The bosonic 
one-body density matrix is given by 
\beq 
n({\bf r},{\bf r}') = 
\langle {\hat \phi}^+({\bf r})\, {\hat \phi}({\bf r}') \rangle \; , 
\eeq
where the average $\langle \cdot\cdot\cdot \rangle$ can be a thermal average 
or a zero-temperature average. Its diagonal part is the average local 
density of bosons, i.e. 
$n({\bf r}) = n({\bf r},{\bf r}) = 
\langle {\hat \phi}^+({\bf r}) \, {\hat \phi}({\bf r}) \rangle$, 
while the average total number of bosons reads 
\beq 
N = \int \langle {\hat \phi}^+({\bf r}) \, {\hat \phi}({\bf r}) \rangle 
\, d^3{\bf r} \; .  
\eeq 
As previously discussed, Penrose and Onsager \cite{penrose} used 
the ODLRO of the one-body density matrix of a uniform bosonic system 
to determine the condensate number $N_0$ of bosons. 
For a non-uniform bosonic system 
this condensate number $N_0$ is nothing else than 
the largest eigenvalue of the one-body density matrix \cite{leggett}. 
As shown by Yukalov \cite{yukalov}, in the thermodynamic limit 
this is equivalent to the spontaneous symmetry breaking 
of $U(1)$ gauge symmetry, which gives 
\beq 
N_0 = \int |\langle {\hat \phi}({\bf r}) \rangle |^2 \, d^3{\bf r} \; 
\eeq
as the average number of condensed bosons. 

A quantum system of interacting identical fermions with two spin components 
($\sigma=\uparrow,\downarrow$) can be described 
by the fermionic field operator ${\hat \psi}_{\sigma}({\bf r})$, 
which satisfies the familiar anti-commutation rules \cite{leggett}
\beq 
\left\{
{\hat \psi}_{\sigma}({\bf r}), {\hat \psi}^+_{\sigma'}({\bf r}') \right\} = 
\delta({\bf r}-{\bf r}')\, \delta_{\sigma,\sigma'} \; , 
\quad\quad 
\left\{ 
{\hat \psi}_{\sigma}({\bf r}), {\hat \psi}_{\sigma'}({\bf r}') 
\right\}  = 
\left\{
{\hat \psi}^+_{\sigma}({\bf r}), {\hat \psi}^+_{\sigma'}({\bf r}')
\right\}  = 0 \; , 
\eeq
where $\left\{{\hat A},{\hat B}\right\}={\hat A}{\hat B}+{\hat B}{\hat A}$ 
and $\delta_{\sigma,\sigma'}$ is the Kronecher symbol. The fermionic 
one-body density matrix is given by 
\beq 
n_{\sigma,\sigma'}({\bf r},{\bf r}') = 
\langle {\hat \psi}^+_{\sigma}({\bf r})\, 
{\hat \psi}_{\sigma'}({\bf r}') \rangle \; .  
\eeq
Its diagonal part is the average local 
density of fermions with spin $\sigma$, i.e. 
$n_{\sigma}({\bf r}) = n_{\sigma,\sigma}({\bf r},{\bf r}) = 
\langle {\hat \psi}^+_{\sigma}({\bf r}) \, 
{\hat \psi}_{\sigma}({\bf r}) \rangle$,  
while the average total number of fermions reads 
\beq 
N = \sum_{\sigma=\uparrow,\downarrow} 
\int \langle {\hat \psi}^+_{\sigma}({\bf r}) \, 
{\hat \psi}_{\sigma}({\bf r}) \rangle 
\, d^3{\bf r} \; .  
\label{def-n}
\eeq 
Yang \cite{yang} suggested that for a uniform strongly-interacting 
fermionic system the number $N_0$ of Bose-condensed fermions, that is 
twice the number of condensed fermionic pairs, is related 
by the ODLRO of the fermionic two-body density matrix, given by 
\beq 
n_{\sigma_1,\sigma_2,\sigma_1',\sigma_2'}
({\bf r}_1,{\bf r}_2,{\bf r}_1',{\bf r}_2') 
= \langle 
{\hat \psi}^+_{\sigma_1}({\bf r}_1) \, 
{\hat \psi}_{\sigma_2}^+({\bf r}_2) \, 
{\hat \psi}_{\sigma_2'}({\bf r}_2') \, 
{\hat \psi}_{\sigma_1'}({\bf r}_1')
\rangle \; .  
\eeq
For a generic (non-uniform) strongly-interacting system of 
identical fermions this condensate number $N_0$ is nothing else than 
twice the largest eigenvalue of the fermionic two-body 
density matrix \cite{leggett}. 
In the thermodynamic limit this is equivalent \cite{campbell} to the 
spontaneous symmetry breaking of $SU(2)$ gauge symmetry, 
which gives 
\beq 
N_0 = 2 \sum_{\sigma,\sigma'=\uparrow,\downarrow} 
\int |\langle {\hat \psi}_{\sigma}({\bf r}) \, 
{\hat \psi}_{\sigma'}({\bf r}') \rangle|^2  
\, d^3{\bf r} \, d^3{\bf r}' \; 
\label{def-n0}
\eeq
as the average number of condensed fermions. It is important to stress that 
within the BCS theory of superconductors and fermionic superfluids 
the condensate number $N_0$ fermions can be much smaller than the 
average total number of fermions $N$. Moreover, in a uniform system 
$N_0$ is twice the average number of fermionic pairs in the (pseudo) 
Bose-Einstein condensate, i.e. the number of fermionic pairs 
which have their center of mass with zero linear momentum. 

\section{Ultracold and dilute atomic gases}

The shifted Hamiltonian of the uniform two-spin-component 
Fermi superfluid made of ultracold atoms is given by  
\beqa 
{\hat H}' =
\int d^3{\bf r} \
\sum_{\sigma=\uparrow , \downarrow}
\ {\hat \psi}^+_{\sigma}({\bf r}) 
\left(-{\hbar^2\over 2 m}\nabla^2- \mu \right) {\hat \psi}_{\sigma}({\bf r}) 
+
g \ {\hat \psi}^+_{\uparrow}({\bf r})\ {\hat \psi}^+_{\downarrow}({\bf r})\  
{\hat \psi}_{\downarrow}({\bf r})\ {\hat \psi}_{\uparrow}({\bf r}) \; , 
\label{ham} 
\eeqa 
where ${\hat \psi}_{\sigma}({\bf r})$ is the field operator 
that annihilates a fermion of spin $\sigma$ 
in the position ${\bf r}$, while ${\hat \psi}_{\sigma}^+({\bf r})$ 
creates a fermion of spin $\sigma$ in ${\bf r}$. Here $g<0$ 
is the strength of the attractive fermion-fermion interaction, 
which is approximated by a contract Fermi pseudo-potential \cite{review}
because for ultracold and dilute gases 
the average distance between atoms is much larger than 
the effective radius of the inter-atomic potential \cite{review,leggett}.   
The ground-state average of the number of fermions is given 
by Eq. (\ref{def-n}). 
This total number $N$ is fixed by the chemical potential $\mu$ 
which appears in Eq. (\ref{ham}).

Within the Bogoliubov approach 
the mean-field Hamiltonian derived from Eq. (\ref{ham}) 
can be diagonalized by using the Bogoliubov-Valatin  
representation of the field operator 
${\hat \psi}_{\sigma}({\bf r})$ 
in terms of the anti-commuting quasi-particle 
Bogoliubov operators ${\hat b}_{{\bf k}\sigma}$ 
with amplitudes $u_{\bf k}$ and $v_{\bf k}$ and 
the quasi-particle energy $E_{\bf k}$. 
In this way one finds familiar expressions 
for these quantities: 
\beq 
E_{\bf k}=\left[({\epsilon}_{\bf k}-\mu)^2 + 
\Delta^2\right]^{1/2}
\eeq
and
\beqa 
u_{\bf k}^2 &=& \left( 1 + ({\epsilon}_{\bf k} 
- \mu)/E_{\bf k} \right)/2
\\
v_{\bf k}^2 &=& \left( 1 - ({\epsilon}_{\bf k} 
- \mu )/E_{\bf k} \right)/2 \; , 
\eeqa 
where $\epsilon_{\bf k}=\hbar^2k^2/(2m)$ is the single-particle 
energy. The parameter $\Delta$ is the pairing gap, 
which satisfies the gap equation 
\beq
-{1\over g} = {1\over \Omega} \sum_{\bf k} {1 \over 2E_{\bf k}} \; ,    
\label{bcsGap} 
\eeq
where $\Omega$ is the volume of the uniform system. 
Notice that this equation is ultraviolet divergent and it must be 
regularized. The equation for the total density $n=N/\Omega$ 
of fermions is obtained from Eq. (\ref{def-n}) as 
\beq 
n = {2\over \Omega}\sum_{\bf k} v_{\bf k}^2 \; . 
\label{enne}
\eeq
Finally, from Eq. (\ref{def-n0}) one finds that 
the condensate density $n_0=N_0/\Omega$ of paired fermions 
is given by \cite{sala-odlro,campbell} 
\beq 
n_{0} = {2\over \Omega} \sum_{\bf k} u_{\bf k}^2 v_{\bf k}^2 \; . 
\label{enne-con}
\eeq 

\begin{figure}
\centerline{\epsfig{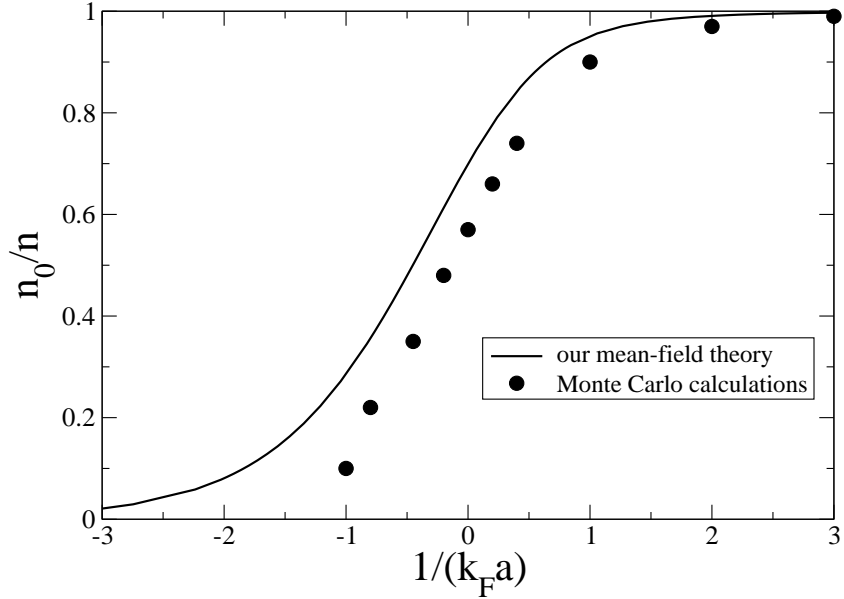}}
\small 
\caption{Condensate fraction of fermionic atoms 
as a function of the inverse interaction strength $1/(k_Fa)$: 
our mean-field theory \cite{sala-odlro} (solid line); 
fixed-node diffusion Monte Carlo results \cite{astra} (filled circles). 
Here $k_F=(3\pi^2n)^{1/3}$ is the Fermi wavenumber, with $n$ the total number 
density of atoms, and $a$ is the s-wave scattering length 
of the inter-atomic potential.} 
\end{figure} 

In three dimensions, a suitable 
regularization \cite{marini} of the gap equation 
is obtained by introducing 
the s-wave scattering length $a$ via the equation 
\beq
-{1\over g} = - {m \over 4 \pi \hbar^2 a} + 
{1 \over \Omega} \sum_{\bf k} \frac{m}{\hbar^2k^2} \,,
\eeq 
and then subtracting this equation from 
the gap equation (\ref{bcsGap}). 
In this way one obtains the three-dimensional regularized gap equation  
\beq 
-{m \over 4 \pi \hbar^2 a} = {1 \over \Omega} 
\sum_{\bf k} \left( { 1 \over 2 E_k} 
- \frac{m}{\hbar^2k^2} 
\right) ,  
\label{bcs3d}  
\eeq 
which can be used to study the full BCS-BEC 
crossover \cite{sala-odlro} by changing the 
amplitude and sign of the s-wave scattering length $a$. 

Taking into account the functional dependence of the
amplitudes $u_k$ and $v_k$ on $\mu$ and $\Delta$, 
one finds \cite{sala-odlro} the very nice formula
\beq 
n_0 = {m^{3/2} \over 8 \pi \hbar^3} \,
\Delta^{3/2} \sqrt{{\mu\over \Delta}+\sqrt{1+{\mu^2 \over \Delta^2} }} \; ,  
\label{godgiven}
\eeq 
which shows the not trivial relationship between 
the energy gap $\Delta$ and the condensate density $n_0$. 
By the same techniques, also the two BCS-BEC equations 
can be written in a more compact form as 
\beq 
-{1\over a} = {2 (2m)^{1/2} \over \pi \hbar^3} \,
\Delta^{1/2} \, I_1\!\left({\mu \over \Delta}\right)  \, , 
\label{gbcs1} 
\eeq 
\beq  
n = {(2m)^{3/2} \over 2 \pi^2 \hbar^3} \,
\Delta^{3/2} \, I_2\!\left({\mu \over \Delta}\right)  \, ,
\label{gbcs2} 
\eeq 
where $I_1(x)$ and $I_2(x)$ are two monotonic 
functions \cite{marini} given by 
\beqa 
I_1(x) &=& \int_0^{+\infty} y^2 
\left( {1\over \sqrt{(y^2-x)^2+1}} - {1\over y^2} \right) dy \; ,
\\
I_2(x) &=& \int_0^{+\infty} y^2 \left( 1 - {y^2-x\over \sqrt{(y^2-x)^2+1}}
\right) dy \; .  
\eeqa

In Fig. 1 we report the condensate fraction $n_0/n$ of fermionic 
atoms in the BCS-BEC crossover as a function of the inverse 
interaction strength $1/(k_Fa)$ obtained with this mean-field theory 
\cite{sala-odlro,ortiz} (solid line). In the figure 
we compare our calculations \cite{sala-odlro} with the 
fixed-node diffusion Monte Carlo results (filled circles) 
obtained by Astrakharchik, Boronat, Casulleras, 
and S. Giorgini \cite{astra} with $N=66$ fermions and 
a tunable square-well potential. Remarkably, the agreement 
between the two theoretical approaches is better 
in the BEC side of the crossover. On the other hand, 
as discussed in Refs. \cite{sala-odlro,furbone}, our Eq. (\ref{godgiven}) 
is in full agreement with the experimental data of 
the MIT group \cite{zwierlein} in the BCS side of the crossover. 

\section{Nuclear matter} 

Let us now consider the nuclear matter, and in particular 
the bulk neutron matter, which is a dense Fermi liquid made of 
two-component (spin up and down) neutrons. 
The shifted Hamiltonian of the uniform neutron matter 
can be written as 
\beqa 
{\hat H}' &=& 
\int d^3{\bf r} \
\sum_{\sigma=\uparrow , \downarrow}
\ {\hat \psi}^+_{\sigma}({\bf r}) 
\left(-{\hbar^2\over 2 m}\nabla^2- \mu \right) {\hat \psi}_{\sigma}({\bf r}) 
\label{mazzabubu}
\\
&+&
\int d^3{\bf r} \ d^3{\bf r}' \  
{\hat \psi}^+_{\uparrow}({\bf r})\ {\hat \psi}^+_{\downarrow}({\bf r}')\ 
V({\bf r}-{\bf r}')\ 
{\hat \psi}_{\downarrow}({\bf r}')\ {\hat \psi}_{\uparrow}({\bf r}) \; , 
\nonumber
\eeqa 
where ${\hat \psi}_{\sigma}({\bf r})$ is the field operator 
that annihilates a neutron of spin $\sigma$ 
in the position ${\bf r}$, while ${\hat \psi}_{\sigma}^+({\bf r})$ 
creates a neutron of spin $\sigma$ in ${\bf r}$. Here $V({\bf r}-{\bf r}')$ 
is the neutron-neutron potential characterized by 
s-wave scattering length $a=-18.5$ fm and effective 
range $r_e=2.7$ fm \cite{matsuo}. 

One can apply the familiar Bogoliubov approach to diagonalize 
the mean-field quadratic Hamiltonian derived from Eq. (\ref{mazzabubu}), 
but now the paring gap  $\Delta_{\bf k}$ 
depends explicitly on the wave number ${\bf k}$ and 
satisfies the integral equation 
\beq
\Delta_{\bf q} = \sum_{\bf k} V_{{\bf q}{\bf k}}  
\ {\Delta_{\bf k} \over 2E_{\bf k}} \; ,  
\eeq 
where 
$V_{{\bf q}{\bf k}}=
\langle {\bf q},-{\bf q}|V|{\bf k},-{\bf k}\rangle$ 
is the wave-number representation of the neutron-neutron potential, and 
\beq 
E_{\bf k} = \sqrt{\left({\hbar^2k^2\over 2m}-\mu\right)^2
+|\Delta_{\bf k}|^2} \; . 
\eeq 
Under the simplifying assumptions $\mu \simeq \epsilon_F = {\hbar^2\over 2m} 
(3\pi^2 n)^{2/3}$ and $\Delta_{\bf k} \simeq \Delta$, 
in the continuum limit we determine the condensate 
fraction as \cite{sala-neutron}
\beq 
{n_0\over n} = {\pi\over 2^{5/2}} {\sqrt{{\epsilon_F\over \Delta}
+\sqrt{1+{\epsilon_F^2\over\Delta^2}}}\over I_2({\epsilon_F\over \Delta})} 
\label{frac}
\eeq 
Notice that in the deep BCS regime where $\Delta/\epsilon_F \ll 1$ one finds 
\beq 
{n_0\over n} = {3\pi\over 8} {\Delta\over \epsilon_F} \; . 
\label{frac-bcs}
\eeq

\begin{figure}
\centerline{\epsfig{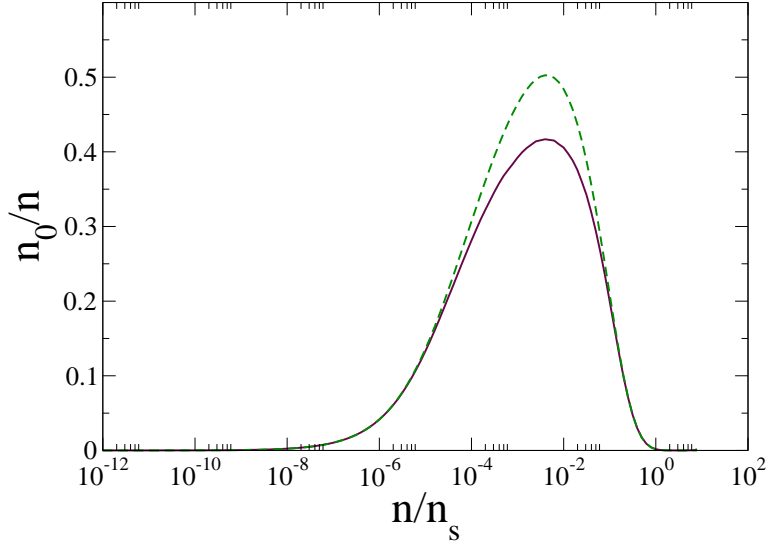}}
\small 
\caption{Condensate fraction $n_{0}/n$ 
of neutron pairs in neutron matter as a function of the scaled 
neutron number density $n/n_s$, 
where $n_s=0.16$ fm$^{-3}$ is the nuclear saturation density 
(see also \cite{sala-neutron}). 
The solid line is obtained by using Eqs. (\ref{frac}) and (\ref{genius}). 
The dashed line is obtained by using Eqs. (\ref{frac-bcs}) and 
(\ref{genius}).}
\end{figure} 

Fitting the numerical data of $\Delta/\epsilon_F$ vs $k_F$ 
obtained by Matsuo \cite{matsuo} from realistic neutron-neutron 
potentials we get the formula 
\beq 
{\Delta\over \epsilon_F} = 
{\beta_0 k_F^{\beta_1} \over \exp({k_F^{\beta_2}/\beta_3)}-\beta_3}
\label{genius}
\eeq
with the following fitting parameters: 
$\beta_0=2.851$, $\beta_1=1.942$, $\beta_2=1.672$,
$\beta_3=0.276$, $\beta_4=0.975$. 
By using this fitting formula and Eq. (\ref{frac}) 
we finally get the condensate fraction of neutron matter as a function 
of the neutron density $n$ \cite{sala-neutron}.

The condensate fraction $n_0/n$ of neutron pairs is shown 
in Fig. 2 as a function of the scaled density $n/n_s$, 
where $n_s=0.16$ fm$^{-3}$ is the nuclear saturation density. Notice that 
the horizontal axis is in logarithmic scale. 
At very low neutron density $n$ the neutron matter behaves like a quasi-ideal 
Fermi gas with weakly correlated Cooper pairs 
and the condensate fraction $n_{0}/n$ is 
exponentially small. By increasing the neutron density 
$n$ the attractive tail of the neutron-neutron potential becomes 
relevant and the condensate fraction $n_{0}/n$ grows significantly.  
The maximum of the condensate 
fraction is $(n_{0}/n)_{max}= 0.42$ 
at the neutron density $n=5.3\cdot 10^{-4}$ fm$^{-3}$ which corresponds to 
the Fermi wave number $k_F=(3\pi^2n)^{1/3} = 0.25$ fm$^{-1}$. 
By further increasing 
the density $n$ the repulsive core of the neutron-neutron potential 
plays an important role in destroying the correlation of Cooper pairs 
and the condensate fraction $n_{0}/n$ slowly goes to zero. 
Remarkably, the results 
of Fig. 2 are fully consistent with the Monte Carlo 
value $n_{0}/n \simeq 0.35$ at $n=0.003$ fm$^{-3}$ 
one extracts from the finite-temperature Path Integral 
Monte Carlo data of Wlazlowski and Magierski \cite{magierski1,magierski2}. 

\section{Neutron stars}

Neutron stars are astronomical compact objects which can result 
from the gravitational collapse of a massive star during a supernova event. 
Such stars are mainly composed of neutrons. 
Neutron stars are very hot and are supported against 
further collapse by Fermi pressure. 
A typical neutron star has a mass $M$ between $1.35$ and about $2.0$  
solar masses with a corresponding radius $R$ of about $12$ km. 
Notice that in the crust of neutron stars one 
estimates \cite{zane} a temperature $T\simeq 10^8$ K, 
while the critical temperature of the normal-superfluid transition is 
$T_c\simeq 10^{10}$ K. Thus the crust of neutron stars is superfluid. 

\begin{figure}
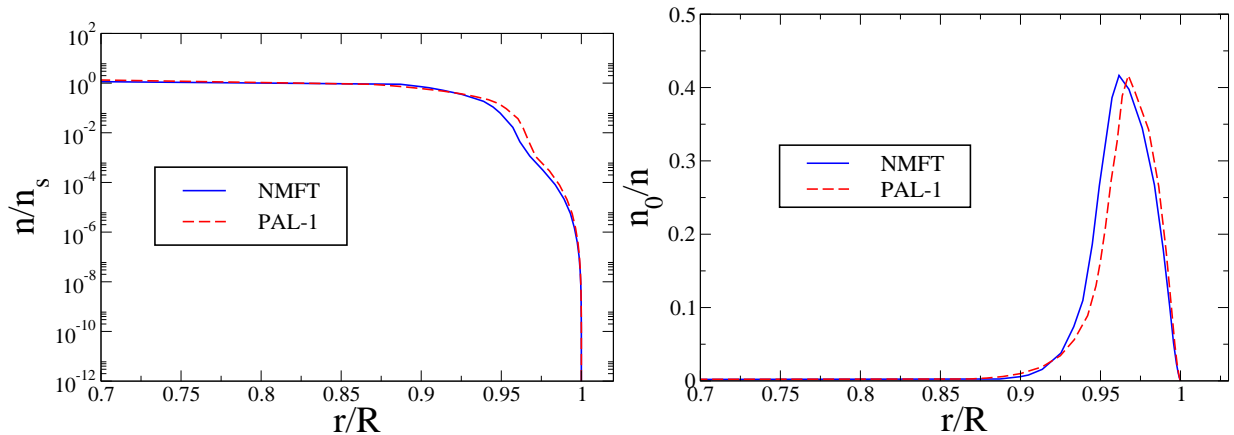

%\centerline
{\epsfig{file=highden-f4a.eps,width=8.cm,clip=}}
%\centerline
{\epsfig{file=highden-f4b.eps,width=8.cm,clip=}}
\small
\caption{$1.4$ solar mass neutron star. 
Left panel: Scaled density profile $n/n_s$ vs scaled distance $r/R$. 
Here $n_s=0.16$ fm$^{-3}$ is the nuclear 
saturation density and $R$ is the radius of the star. 
Right panel: condensate fraction $n_0/n$ of neutron pairs 
vs scaled distance $r/R$. 
Solid lines are obtained with the BCS model of bulk 
neutron matter \cite{walecka}. 
Dashed lines are obtained with the more sophisticated model 1 of 
Prakash, Ainsworth, and Lattimer \cite{lattimer}.} 
\end{figure} 

Some years ago, Datta, Thampan, and Bhattacharya \cite{datta} 
have calculated several mass density profiles $\rho(r)$ 
of spherical and non-rotating neutron stars by solving the Tolman-
Oppenheimer-Volkoff (TOV) equation, which 
describes the interplay between the expulsive kinetic pressure of the star 
and its gravitational self-attraction \cite{star,zane}. 
Datta, Thampan, and Bhattacharya \cite{datta} have solved the TOV equation 
by using various equations of state (EOS) of the nuclear matter. 
In the left panel of Fig. 3 we plot their results in the case of $1.4$ 
solar mass neutron star. In particular, we report 
the scaled density profile $n(r)/n_s$ 
of the neutron star as a function of 
the scaled distance $r/R$ from the 
center of the star, where $n_s$ is the nuclear 
saturation density and $R$ is the radius of the star. 
Notice that $n(r)=\rho(r)/m_N$ with $m_N$ the neutron mass. 
The solid line is obtained \cite{datta} solving the TOV equation with 
the simple nuclear EOS of Walecka \cite{walecka}, 
while the dashed line is obtained \cite{datta} solving the TOV equation with 
a more sophisticated EOS, called model 1, 
of Prakash, Ainsworth and Lattimer \cite{datta,lattimer}. 

In the previous section we have found a fitting formula for the 
condensate fraction $n_0/n$ of neutron matter 
as a function of the bulk neutron density $n$. 
Knowing the density profile $n(r)$ of a 
neutron star, 
i.e. the neutron density $n$ as a function of the distance $r$ from the 
center of a neutron star, we can determine (local density approximation) 
the condensate fraction $n_0/n$ of the neutron star 
as a function of the distance $r$. 

The results are shown in the right panel of Fig. 3. 
The figure shows that the two EOS give very similar results and 
that a relevant condensate fraction 
appears only in the crust of the neutron star 
in the region between $r/R=0.85$ and $r/R=1$ 
with a maximum value $\simeq 0.4$ 
at $r/R\simeq 0.96$. We stress that this suggestive 
plot can be certainly improved because in neutron stars 
the hadronic matter is not made only of neutrons \cite{star,zane}. 
On the other hand, it is exactly in the crust of neutron stars 
that it is expected to find the dilute neutron matter we have 
considered to derive Eq. (\ref{frac}). 

\section{Conclusions}

We have seen that the condensate fraction of Cooper pairs 
can be calculated in various superfluid fermionic systems: 
dilute atomic gases, dense neutron matter and neutron stars. 
We observe that, while the condensate fraction 
in ultracold gases of fermionic atoms has been measured in two 
sophisticated experiments by measuring the momentum distribution 
of pairs \cite{zwierlein,ueda}, it remains open the exciting problem 
of finding reliable observational signatures of 
the condensate fraction of neutron-neutron pairs 
in atomic nuclei and in neutron stars. 
In conclusion, we point out that recently we have studied the 
behavior of the condensate fraction in ultracold and dilute 
gases of fermionic atoms not only in the 3D uniform system but also 
in other configurations: 2D uniform system \cite{sala-2d}, 
2D system on a square lattice \cite{sala-2dl}, 3D and 2D 
uniform system with spin-orbit coupling \cite{sala-so}, 
3D and 2D uniform system with three-spin components \cite{sala-3spin}, 
and 3D uniform system with a narrow Feschbach 
resonance \cite{sala-narrow}. 

\ack

The author thanks Masayuki Matsuo for making available his 
numerical data and Roberto Turolla for useful discussions. 
The author acknowledges research grants from 
Universit\`a di Padova (Progetto di Ricerca di Ateneo 2012-201),  
Fondazione CARIPARO (Progetto di Eccellenza 2012-2013), 
and Ministero dell'Istruzione Universit\`a e Ricerca 
(Progetto PRIN call 2011-2012).

\end{document}